# Relativistic equilibrium velocity distribution, nuclear fusion reaction rate and the solar neutrino problem

Jian-Miin Liu*
Department of Physics, Nanjing University
Nanjing, The People's Republic of China
*On leave. E-mail address: liu@phys.uri.edu

ABSTRACT
   In solar interior, it is the equilibrium velocity distribution of few high-energy protons and nuclei that participates in determining nuclear fusion reaction rates. So, it is inappropriate to use the Maxwellian velocity distribution to calculate the rates of solar nuclear fusion reactions. We have to use the relativistic equilibrium velocity distribution for the purpose. The nuclear fusion reaction rate based on the relativistic equilibrium velocity distribution has a reduction factor with respect to that based on the Maxwellian distribution. The reduction factor depends on the temperature, reduced mass and atomic numbers of the studied nuclear fusion reactions, in other words, it varies with the sort of neutrinos. Substituting the relativistic equilibrium velocity distribution for the Maxwellian distribution is not important for the calculation of solar sound speeds. The relativistic equilibrium velocity distribution, if adopted in standard solar models, will lower solar neutrino fluxes and change solar neutrino energy spectra but maintain solar sound speeds. This velocity distribution is possibly a solution to the solar neutrino problem.

1. INTRODUCTION
   The solar neutrino problem is a long-standing puzzle in modern physics. Besides the difficulties in understanding the observed data of the $^8$B-, $^7$Be-, CNO-, pp- and pep-neutrino fluxes in different experiments from the viewpoint of solar neutrino energy spectra, it contains discrepancies between the measured solar neutrino fluxes and those predicted by standard solar models [1-16]. Since standard solar models lead to a very close agreement about sound speeds between theoretical calculations and helioseimological observations [17], the difficulties and the discrepancies are regarded as an evidence for new physics. Massive neutrinos, neutrino flavor oscillations and lepton flavor non-conservation beyond standard electroweak model have been suggested for this new physics [18]. However, this kind of new physics is still waiting to be confirmed, especially out of solar neutrino arena.
   In this paper, we suggest another kind of new physics, for standard solar models, which concerns the relativistic generalization of the Maxwellian velocity distribution. Based on the relativistic equilibrium velocity distribution, we derive a new formula for nuclear fusion reaction rate. This formula, once adopted in standard solar models, will lower solar neutrino fluxes and change solar neutrino energy spectra but maintain solar sound speeds. The paper consists of seven sections: introduction, a personal opinion on standard solar models, relativistic velocity space, relativistic equilibrium velocity distribution, nuclear fusion reaction rate, a possible solution to the solar neutrino problem, concluding remarks.

2. A PERSONAL OPINION ON STANDARD SOLAR MODELS
   Standard solar models [10-16] deal with our Sun as a typical one of main sequence stars in stellar evolution. Depending on several initial inputs, these models successfully describe the Sun's properties such as its total mass, radius, luminosity, temperature and compositions of heavy elements at the Sun age. Some models with the diffusion mechanism about helium and heavy elements achieve further success in being very consistent with the observed helioseimological data. But, all these models predict solar neutrino signals unsuccessfully.
   All standard solar models take

$$R_M = \frac{1}{(1+\delta_{12})} N_1 N_2 \frac{4c}{\sqrt{3}K_B T} S_{eff} \exp[-\lambda] (2\pi z_1 z_2 \frac{K_B T}{\mu c^2} \frac{e^2}{\hbar c})^{1/3}, \qquad (1)$$

as a formula for the input parameters of solar nuclear fusion reaction rates. This formula is based on the Maxwellian velocity distribution. Maxwellian velocity distribution is non-relativistic.

Solar core is a dense plasma of high temperature $T_c=14.9 \times 10^6$ K and high density $\rho_c=150 \text{g/cm}^3$. Solar core produces its neutrinos primarily through nuclear fusion reactions in the proton-proton cycle and the CNO cycle. In solar interior, to create a nuclear fusion reaction, a proton or nucleus must penetrate the repulsive Coulomb barrier and be close to another proton or nucleus so that the strong interaction between them acts. As the height of Coulomb barrier is far above solar thermal energy $K_B T_c$, their ratio is typically greater than a thousand [16,19], solar nuclear fusion reactions can occur only among few high-energy protons and nuclei. On the other hand, as the interacting protons and nuclei, at the temperatures and densities in solar interior, reach their equilibrium distribution in the period of time that is infinitesimal compared to the mean lifetime of a nuclear fusion reaction [16,19], an equilibrium velocity distribution is quite applicable to the calculation of the rates of solar nuclear fusion reactions. In solar interior, therefore, it is the equilibrium velocity distribution of these few high-energy protons and nuclei that participates in determining solar nuclear fusion reaction rates, and hence, solar neutrino fluxes and solar neutrino energy spectra. In the situation, it is inappropriate to use the Maxwellian velocity distribution to calculate the rates of solar nuclear fusion reactions. We have to use the relativistic equilibrium velocity distribution for the purpose. Fig.1 sketches the situation. Solid line is the Maxwellian velocity distribution while dash line is the relativistic equilibrium velocity distribution which fits to the Maxwellian for low-energy particles but substantially differs from the Maxwellian for high-energy particles. Low-energy protons or nuclei in shadow area do not take part in solar nuclear fusion reactions. The difference between the Maxwellian velocity distribution and the relativistic equilibrium velocity distribution for high-energy protons or nuclei in unshaded area becomes non-negligible when we calculate the rates of solar nuclear fusion reactions.

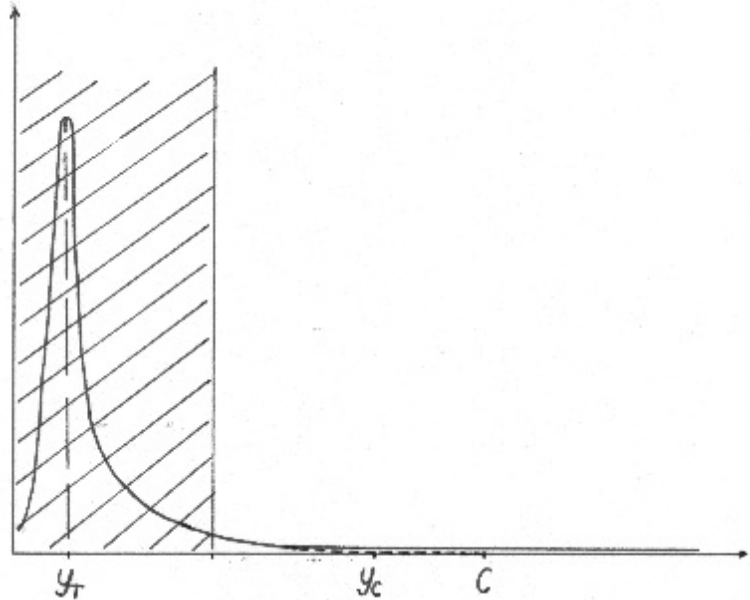

Fig.1   The Maxwellian velocity distribution (solid line) and the relativistic equilibrium velocity distribution (dash line) under the conditions in the Sun. The relativistic equilibrium velocity distribution fits to the Maxwellian for low-energy particles but substantially differs from the Maxwellian for high-energy particles. It falls off to zero as y goes to c. $y_T$ and $y_C$ respectively correspond to solar thermal energy and the Coulumb barrier. Coulumb barrier is far above solar thermal energy, their ratio is typically greater than a thousand. As low-energy protons or nuclei in shadow area do not take part in solar nuclear fusion reactions, the difference between the Maxwellian velocity distribution and the relativistic equilibrium velocity distribution for high-energy protons or nuclei in unshaded area becomes non-negligible for the calculation of the rates of solar nuclear fusion reactions.

The formula in Eq.(1) is not suited to the calculation of the rates of solar nuclear fusion reactions. Taking this formula, in our opinion, standard solar models carry some errors in their input parameters of solar nuclear fusion reaction rates.

3. RELATIVISTIC VELOCITY SPACE

Velocity space is a space in which pairs of points represent relative velocities. In the pre-relativistic mechanics, the velocity space is

$$dY^2 = \delta_{rs} dy^r dy^s, \quad r,s=1,2,3, \tag{2}$$

in the usual (Newtonian) velocity-coordinates $\{y^r\}$, r=1,2,3, where $y^r$ is the well-defined Newtonian velocity. This velocity space is characterized by unboundedness and the Galilean addition law in the usual velocity-coordinates. The relativistic velocity space is [20,21,27]

$$dY^2 = H_{rs}(y) dy^r dy^s, \quad r,s=1,2,3, \tag{3a}$$
$$H_{rs}(y) = c^2 \delta^{rs}/(c^2-y^2) + c^2 y^r y^s/(c^2-y^2)^2, \quad \text{real } y^r \text{ and } y<c, \tag{3b}$$

in the usual velocity-coordinates or

$$dY^2 = \delta_{rs} dy'^r dy'^s, \quad r,s=1,2,3, \tag{4}$$

in the so-called primed velocity-coordinates $\{y'^r\}$, r=1,2,3, where $y=(y^r y^r)^{1/2}$, c is the Newtonian speed of light, $y'^r$ is named the primed velocity.

In the relativistic velocity space, the usual and the primed velocity-coordinates are in connection by

$$dy'^r = A^r_s(y) dy^s, \quad r,s=1,2,3, \tag{5a}$$
$$A^r_s(y) = \gamma \delta^{rs} + \gamma(\gamma-1) y^r y^s / y^2, \tag{5b}$$

where

$$\gamma = 1/(1-y^2/c^2)^{1/2}, \tag{6}$$

because

$$H_{rs}(y) = \delta_{pq} A^p_r(y) A^q_s(y), \quad p,q,r,s=1,2,3. \tag{7}$$

Using the calculation techniques in Riemann geometry, at some length, we find from Eqs.(3a-3b),

$$Y(y_1^r, y_2^r) = \frac{c}{2} \ell n \frac{b+a}{b-a} \tag{8a}$$

with

$$b = c^2 - y_2^r y_1^r, \quad r=1,2,3, \tag{8b}$$
$$a = \{(c^2 - y_1^i y_1^i)(y_2^j - y_1^j)(y_2^j - y_1^j) + [y_1^k(y_2^k - y_1^k)]^2\}^{1/2}, \quad i,j,k=1,2,3, \tag{8c}$$

while we find from Eq.(4),

$$Y(y'^r_1, y'^r_2) = [(y'^r_2 - y'^r_1)(y'^r_2 - y'^r_1)]^{1/2}. \tag{9}$$

In Eqs.(8a-8c) and (9), we can call $y_1^r=0$, $y_2^r=y^r$, $y'^r_1=0$ and $y'^r_2=y'^r$ to obtain the relationship between a Newtonian velocity $y^r$ and its corresponding primed velocity $y'^r$,

$$y'^r = [\frac{c}{2y} \ell n \frac{c+y}{c-y}] y^r, \quad r=1,2,3, \tag{10a}$$

$$y' = \frac{c}{2} \ell n \frac{c+y}{c-y}, \tag{10b}$$

where $(y^1, y^2, y^3)$ and $(y'^1, y'^2, y'^3)$ represent the same point in the relativistic velocity space, $y'=(y'^r y'^r)^{1/2}$, r=1,2,3. The relativistic velocity space is characterized by unboundedness and the Galilean addition law in the primed velocity-coordinates and by a finite boundary at y=c and the Einstein addition law in the usual velocity-coordinates. It has turned out that the Galilean addition law of primed velocities links up with the Einstein addition law of Newtonian velocities.

To seek for the relationship between a relative Newtonian velocity and its corresponding primed velocity, we let particle 1 and particle 2 move with Newtonian velocities $y_1^r$ and $y_2^r$, r=1,2,3, respectively and denote the relative Newtonian velocity of particle 2 to particle 1 with $v^r$, r=1,2,3. If three corresponding primed velocities are respectively $y_1'^r$, $y_2'^r$, $v'^r$, r=1,2,3, the Galilean addition law among them reads

$$v'^r = y'^r_2 - y'^r_1, \quad r=1,2,3, \tag{11}$$

and the Einstein addition law among the three Newtonian velocities is

$$v^r = \sqrt{1 - y_1^2/c^2}\ \{(y_2^r - y_1^r) + (\frac{1}{\sqrt{1-y_1^2/c^2}} - 1) y_1^r \frac{y_1^s(y_2^s - y_1^s)}{y_1^2}\}/[1 - \frac{y_2^k y_1^k}{c^2}], \ r,s,k=1,2,3.$$

(12)

On the other hand, Eqs. (8a) and (9) give us

$$[(y'_2{}^r - y'_1{}^r)(y'_2{}^r - y'_1{}^r)]^{1/2} = \frac{c}{2}\ell n \frac{b+a}{b-a},$$

or equivalently

$$c^2 \tanh^2\{[(y'_2{}^r - y'_1{}^r)(y'_2{}^r - y'_1{}^r)]^{1/2}/c\} = c^2\{(c^2 - y_1^i y_1^i)(y_2^j - y_1^j)(y_2^j - y_1^j) + [y_1^k(y_2^k - y_1^k)]^2\}/(c^2 - y_1^r y_1^r)^2,\quad (13)$$

where two expressions $b$ and $a$ are respectively in Eqs.(8b) and (8c). We separately apply Eqs.(11) and (12) to the left-hand and the right-hand sides of Eq.(13), and get

$$c^2 \tanh^2\{v'/c\} = v^2,\tag{14}$$

where $v' = (v'^r v'^r)^{1/2}$ and $v = (v^r v^r)^{1/2}$. Eq.(14) implies

$$v'^r = (\frac{c}{2v}\ell n \frac{c+v}{c-v}) v^r, \ r=1,2,3,\tag{15}$$

$$v' = \frac{c}{2}\ell n \frac{c+v}{c-v},\tag{16}$$

which specify the relationship between the relative Newtonian velocity $v^r$ and its corresponding primed velocity $v'^r$. Differentiating Eq.(16) immediately yields

$$dv' = \frac{dv}{(1-v^2/c^2)}.\tag{17}$$

## 4. RELATIVISTIC EQUILIBRIUM VELOCITY DISTRIBUTION

Relying on two assumptions, (1) the velocity distribution function is spherically symmetric and (2) the x-, y- and z- components of velocity are statistically independent, Maxwell derived his equilibrium velocity distribution [22],

$$M(y^1,y^2,y^3) dy^1 dy^2 dy^3 = N(\frac{m}{2\pi K_B T})^{3/2} \exp[-\frac{m}{2K_B T}(y)^2] dy^1 dy^2 dy^3,\tag{18a}$$

$$M(y)dy = 4\pi N(\frac{m}{2\pi K_B T})^{3/2} (y)^2 \exp[-\frac{m}{2K_B T}(y)^2] dy,\tag{18b}$$

where N is the number of particles, m their rest mass, T the temperature and $K_B$ the Boltzmann constant. These assumptions reflect structural characteristics of the pre-relativistic velocity space in Eq.(2).

The Euclidean structure of the rlativistic velocity space in the primed velocity-coordinates convinces us of the Maxwellian distribution of primed velocities,

$$P(y'^1,y'^2,y'^3) dy'^1 dy'^2 dy'^3 = N(\frac{m}{2\pi K_B T})^{3/2} \exp[-\frac{m}{2K_B T}(y')^2] dy'^1 dy'^2 dy'^3 \tag{19a}$$

and

$$P(y')dy' = 4\pi N(\frac{m}{2\pi K_B T})^{3/2} (y')^2 \exp[-\frac{m}{2K_B T}(y')^2] dy'.\tag{19b}$$

The relativistic equilibrium distribution of Newtonian velocities will be given by using Eq.(10b) and

$$dy'^1 dy'^2 dy'^3 = \gamma^4 dy^1 dy^2 dy^3$$

and

$$dy' = \gamma^2 dy,$$

which are respectively inferred from Eqs.(5a-5b) and Eq.(10b), in Eqs.(19a-19b):

$$P(y^1,y^2,y^3) dy^1 dy^2 dy^3 = N \frac{(m/2\pi K_B T)^{3/2}}{(1-y^2/c^2)^2} \exp[-\frac{mc^2}{8K_B T}(\ell n \frac{c+y}{c-y})^2] dy^1 dy^2 dy^3,\tag{20a}$$

$$P(y)dy = \pi c^2 N \frac{(m/2\pi K_B T)^{3/2}}{(1-y^2/c^2)} (\ell n \frac{c+y}{c-y})^2 \exp[-\frac{mc^2}{8K_B T}(\ell n \frac{c+y}{c-y})^2] dy. \tag{20b}$$

Distribution functions $P(y^1,y^2,y^3)$ and $P(y)$ respectively reduce to $M(y^1,y^2,y^3)$ and $M(y)$ for small velocities, $y/c \ll 1$. Substantially differing from to $M(y^1,y^2,y^3)$ and $M(y)$, they fall off to zero as y goes to c slower than any exponential decay, $\exp\{-[2c/(c-y)]^B\}$, and faster than any power-law decay, $(c-y)^n$, where B and n are two positive numbers [23]. The relativistic equilibrium velocity distribution has been used to explain the observed non-Maxwellian decay mode of high-energy tails in velocity distributions of astrophysical plasma particles [23].

## 5. NUCLEAR FUSION REACTION RATE

Now let us consider a kind of nuclear fusion reactions taking place between protons or nuclei of type 1 and protons or nuclei of type 2, where protons or nuclei of type i have density $N_i$, mass $m_i$ and atomic number $z_i$, i=1,2. The rate of the considered nuclear fusion reactions is [24]

$$R = \frac{1}{(1+\delta_{12})} N_1 N_2 \langle v\sigma(v) \rangle = \frac{1}{(1+\delta_{12})} N_1 N_2 \int_0^c v\sigma(v) f(v) dv, \tag{21}$$

where v denotes the radial relative velocities of the type 2 protons or nuclei to the type 1 ones, $\sigma(v)$ is a cross section of nuclear fusion reactions of the kind, $\langle \ \rangle$ means the thermodynamic-equilibrium average over v, $0 \leq v < c$, f(v) is the normalized equilibrium distribution, $\int_0^c f(v) dv = 1$.

It has been proved [25] that when velocities of the type 1 particles, as well as those of the type 2 particles, obey the Maxwellian distribution, the relative velocities of the type 2 particles to the type 1 particles obey the like, in which provided we take reduced mass

$\mu = m_1 m_2 / (m_1 + m_2)$.

So, in accordance with Eqs.(19a-19b), the equilibrium distribution for the relative primed velocities of the type 2 particles to the type 1 particles is

$$P^*(v'^1,v'^2,v'^3) dv'^1 dv'^2 dv'^3 = N(\frac{\mu}{2\pi K_B T})^{3/2} \exp[-\frac{\mu}{2K_B T}(v')^2] dv'^1 dv'^2 dv'^3, \tag{22a}$$

$$P^*(v') dv' = 4\pi N(\frac{\mu}{2\pi K_B T})^{3/2} (v'^2) \exp[-\frac{\mu}{2K_B T}(v'^2)] dv'. \tag{22b}$$

Inserting Eqs.(16) and (17) in Eq.(22b), we obtain the relativistic equilibrium distribution for radial relative Newtonian velocities of the type 2 particles to the type 1 particles,

$$P^*(v) dv = \pi c^2 N \frac{(\mu/2\pi K_B T)^{3/2}}{(1-v^2/c^2)} (\ell n \frac{c+v}{c-v})^2 \exp[-\frac{\mu c^2}{8K_B T}(\ell n \frac{c+v}{c-v})^2] dv. \tag{23}$$

Nuclear cross section is a product of three factors: the penetration probability factor, the slowly varying factor S and the factor of the squared de Broglie wavelength. The penetration probability factor is a function of radial relative velocity v. It describes a distribution of the probability for an incoming proton or nucleus with atomic number $z_1$ to penetrate through the repulsive Coulomb barrier of a target proton or nucleus with atomic number $z_2$ due to quantum tunnel effect [26], as well as $P^*(v)/N$ in Eq.(23) is a distribution of the probability for the incoming proton or nucleus to have radial velocity v relative to the target proton or nucleus. We ought to get the penetration probability factor in the same way that we did for $P^*(v)/N$. In other words, we keep $\exp[-\frac{2\pi z_1 z_2 e^2}{\hbar v'}]$ in the primed velocity-coordinates for the penetration probability factor and use Eq.(16) in it to find $\exp[-2\pi z_1 z_2 e^2/\hbar (\frac{c}{2} \ell n \frac{c+v}{c-v})]$ in the usual velocity-coordinates for this factor. The other two factors are together with the penetration probability factor to form the cross section as a whole, so we deal with these two factors consistently, also keeping their original forms in the primed velocity-coordinates and finding their forms in the usual velocity-coordinates with the aid of Eq.(16). Totally, for the cross section, we have

$$\frac{S}{\mu v'^2/2} \exp[-\frac{2\pi z_1 z_2 e^2}{\hbar v'}] \qquad (24)$$

in the primed velocity-coordinates and

$$\{S/\frac{\mu}{2}(\frac{c}{2}\ln\frac{c+v}{c-v})^2\}\exp[-2\pi z_1 z_2 e^2/\hbar (\frac{c}{2}\ln\frac{c+v}{c-v})] \qquad (25)$$

in the usual velocity-coordinates.

In order to calculate nuclear fusion reaction rate R, we take Eq.(25) and P*(v)/N in Eq.(23) and put them into Eq.(21) respectively for $\sigma(v)$ and f(v). Introducing a new variable, $x=\frac{c}{2}\ln\frac{c+v}{c-v}$, we find

$$R = \frac{8\pi}{\mu}(\frac{\mu}{2\pi K_B T})^{3/2}\frac{1}{(1+\delta_{12})}N_1 N_2 \int_0^\infty Sc[\tanh(\frac{x}{c})]\exp[-\frac{\mu x^2}{2K_B T}-\frac{2\pi z_1 z_2 e^2}{\hbar x}]dx. \qquad (26)$$

Calling

$$\tanh(\frac{x}{c}) = \sum_{n=1}^\infty (-1)^{n+1}\frac{2^{2n}(2^{2n}-1)B_n}{(2n)!}(\frac{x}{c})^{2n-1},$$

in a region, where $B_n$, n=1,2,3,------, are the Bernoulli numbers, $B_1=1/6$, $B_2=1/30$, $B_3=1/42$, ------, and using the method of the steepest descents, we further find

$$R = \sum_{n=1}^\infty R_n, \qquad (27)$$

$$R_n = \frac{1}{(1+\delta_{12})}N_1 N_2 \frac{4c}{\sqrt{3K_B T}}S_{eff}\exp[-\lambda](-1)^{n+1}\frac{2^{2n}(2^{2n}-1)B_n}{(2n)!}(2\pi z_1 z_2 \frac{K_B T}{\mu c^2}\frac{e^2}{\hbar c})^{\frac{2n-1}{3}}, \qquad (28)$$

where $S_{eff}$ is the average of slowly varying function S and

$$\lambda = \frac{3}{K_B T}(\frac{\sqrt{\mu K_B T}\pi z_1 z_2 e^2}{\sqrt{2}\hbar})^{2/3}$$

is n-independent. The most effective energy is

$$E_0 = (\frac{\sqrt{\mu K_B T}\pi z_1 z_2 e^2}{\sqrt{2}\hbar})^{2/3},$$

which is also n-independent. In a compact form, we can rewrite R as

$$R = \frac{1}{(1+\delta_{12})}N_1 N_2 \frac{4c}{\sqrt{3K_B T}}S_{eff}\exp[-\lambda]\tanh\{(2\pi z_1 z_2 \frac{K_B T}{\mu c^2}\frac{e^2}{\hbar c})^{1/3}\} \qquad (29)$$

or

$$R = \frac{\tanh Q}{Q}R_M, \qquad (30a)$$

$$Q = (2\pi z_1 z_2 \frac{K_B T}{\mu c^2}\frac{e^2}{\hbar c})^{1/3}, \qquad (30b)$$

where $R_M$ is in Eq.(1).

6. A POSSIBLE SOLUTION TO THE SOLAR NEUTRINO PROBLEM

The nuclear fusion reaction rate based on the relativistic equilibrium velocity distribution has a reduction factor with respect to that based on the Maxwellian velocity distribution: $\tanh Q/Q$. Since $0<Q<\infty$, the reduction factor satisfies $0<\tanh Q/Q<1$. That gives rise to

$$0<R<R_M.$$

The reduction factor depends on the temperature, reduced mass and atomic numbers of the studied nuclear fusion reactions, in other words, it varies with the kind of neutrinos.

The new formula of nuclear fusion reaction rate, Eq.(29) or Eqs.(30a-30b), provides with the lower rates for solar nuclear fusion reactions than the formula Eq.(1). It will hence lower solar neutrino fluxes and change solar neutrino energy spectra in the manner varying with the kind of neutrinos.

The calculation of solar sound speeds, it is associated with all solar ions of low-energy and high-energy. Since most solar ions crowd in the low-energy region even at temperatures and densities in the Sun, since these most low-energy ions are involved in the calculation of solar sound speeds, substituting the relativistic equilibrium velocity distribution for the Maxwellian velocity distribution is not so important for the calculation of solar sound speeds. No change is needed with the current calculation of solar sound speeds.

In this way, the relativistic equilibrium velocity distribution, if adopted in standard solar models, will lower solar neutrino fluxes and change solar neutrino energy spectra but maintain solar sound speeds. The relativistic equilibrium velocity distribution is possibly a solution to the solar neutrino problem.

## 7. CONCLUDING REMARKS

(1) The relativistic equilibrium velocity distribution is a more true law. It fits to the Maxwellian distribution for low-energy particles but substantially differs from the Maxwellian distribution for high-energy particles.

(2) For a statistical calculation relevant to equilibrium velocity distribution, in two cases we have to substitute the relativistic equilibrium velocity distribution for the Maxwellian velocity distribution. One case is where most particles crowd in the high-energy region. The other case is in which most particles crowd in the low-energy region and these particles in the low-energy region are not involved in the statistical calculation. The calculation of the rates of solar nuclear fusion reactions is of the case. When most particles crowd in the low-energy region and, concurrently, these low-energy particles are involved in the statistical calculation, substituting the relativistic equilibrium velocity distribution for the Maxwellian velocity distribution is negligible. The calculation of solar sound speeds belongs here.

(3) The nuclear fusion reaction rate based on the relativistic equilibrium velocity distribution has a reduction factor with respect to that based on the Maxwellian velocity distribution. The reduction factor depends on the temperature, reduced mass and atomic numbers of the studied nuclear fusion reactions, in other words, it varies with the kind of neutrinos.

(4) Standard solar models contain some errors in their input parameters of solar nuclear fusion reaction rates. By taking the relativistic equilibrium velocity distribution and its induced formula of the nuclear fusion reaction rate, standard solar models can be improved. The relativistic equilibrium velocity distribution is possibly a key to resolve the solar neutrino problem.

ACKNOWLEDGMENT
The author greatly appreciates the teachings of Prof. Wo-Te Shen. The author thanks Prof. S. S. Bandola and Dr. P. Rucker for their useful suggestions.

REFERENCES
[1]  R. Davis, D. S. Harmer and K. C. Hoffman, Phys. Rev. Lett., 20, 1205 (1968)
[2]  W. Hampel et al (The Gallex collaboration), Phys. Lett., B447, 127 (1999)
[3]  J. N. Abdurashitov et al (The Sage collaboration), astro-ph/0204245
[4]  V. Gavrin et al (The Sage collaboration), in Proceedings of the 17th International Conference on Neutrino Physics and Astrophysics, eds. K. Huitu et al, World Scientific, Singapore (1997)
[5]  Y. Fukuda et al (The Super-Kamiokande collaboration), Phys. Rev. Lett., 86, 5651 (2001)
[6]  Y. Fukude et al (The Super-Kamiokande collaboration), hep-ex/0205075
[7]  Q. R. Ahmad et al (The SNO collaboration), Phys. Rev. Lett., 87, 071301 (2001)
[8]  Q. R. Ahmad et al (The SNO collaboration), nucl-ex/0204008

[9]     K. Eguchi et al (The KamLand collaboration), hep-ex/0212021
[10]    J. N. Bahcall and M. H. Pinsonneault, Rev. Mod. Phys., 67, 781 (1995)
[11]    A. Dar and G. Shaviv, Astrophys. J., 468, 933 (1996)
[12]    F. Ciacio, S. Degl'Innocenti and B. Ricci, Astron. Astrophys. (Suppl.), 123, 449 (1997)
[13]    O. Richard et al, Astron. Astrophys., 312, 1000 (1996)
[14]    J. A. Guzik and A. N. Cox, Astrophys. J., 411, 394 (1993)
[15]    C. R. Proffitt, Astrophys. J., 425, 849 (1994)
[16]    S. Turck-Chieze et al, Phys. Rep., 230, 57 (1993)
[17]    J. N. Bahcall et al, astro-ph/0209080
[18]    There are many papers on the topic. Go to hep-ph@xxx.lanl.gov for them.
[19]    J. N. Bahcall, Neutrino Astrophysics, Cambridge University Press, Cambridge (1989)
[20]    Jian-Miin Liu, Chaos, Solitons&Fractals, 12, 1111 (2001)
[21]    Jian-Miin Liu, Chaos, Solitons&Fractals, 12, 2149 (2001) [physics/0108045]
[22]    J. C. Maxwell, Phil. Mag., 19, 19 (1860)
[23]    Jian-Miin Liu, cond-mat/0112084
[24]    Jian-Miin Liu, nucl-th/0210058
[25]    D. D. Clayton, Principle of Stellar Evolution and Nucleosythesis, University of Chicago Press, Chicago (1983)
[26]    G. Gamow, Phys. Rev., 53, 595 (1938)
[27]    Jian-Miin Liu, Localization of Lorentz transformation and its induced local Lorentz invariance, physics/0307041